\documentclass[format=sigconf, timestamp=true, authordraft=false]{acmart}

\def\BibTeX{{\rm B\kern-.05em{\sc i\kern-.025em b}\kern-.08em
    T\kern-.1667em\lower.7ex\hbox{E}\kern-.125emX}}

\usepackage[utf8x]{inputenc}

\usepackage{float}

\usepackage{tikz}
\usetikzlibrary{decorations.pathreplacing}
\usepackage{pgfplots}
\usepackage{graphicx}
\usepackage{subcaption}

\setcopyright{rightsretained}

\acmDOI{}

\acmISBN{XXX-X-XXXXX-XXX-X}

\acmYear{2020}



\newcommand{\code}[1]{\texttt{#1}}

\begin{document}

\title{The ReProVide Query-Sequence Optimization in a Hardware-Accelerated DBMS}
\subtitle{Full Paper}


\author{Lekshmi B.G.}
\orcid{0000-0001-7228-926X}
\affiliation{%
  \institution{Friedrich-Alexander-Universität Erlangen-Nürnberg (FAU)}
  \streetaddress{INF6, Martensstr. 3}
  \postcode{91058}
  \city{Erlangen}
  \country{Germany}
}
\email{lekshmi.bg.nair@fau.de}

\author{Andreas Becher}
\affiliation{%
  \institution{Friedrich-Alexander-Universität Erlangen-Nürnberg (FAU)}
  \streetaddress{INF12, Cauerstr. 11}
  \postcode{91058}
  \city{Erlangen}
  \country{Germany}
}
\email{andreas.becher@fau.de}

\author{Klaus Meyer-Wegener}
\orcid{0000-0002-8102-1019}
\affiliation{%
  \institution{Friedrich-Alexander-Universität Erlangen-Nürnberg (FAU)}
  \streetaddress{INF6, Martensstr. 3}
  \postcode{91058}
  \city{Erlangen}
  \state{Germany}
}
\email{klaus.meyer-wegener@fau.de}

\renewcommand{\shortauthors}{Lekshmi B.G., A. Becher, and K. Meyer-Wegener}

\begin{abstract}
Hardware acceleration of database query processing
can be done with the help of FPGAs.
In particular, they are partially reconfigurable at runtime,
which allows for the runtime adaption of the hardware to a variety of queries.
Reconfiguration itself, however, takes some time.
As the affected area of the FPGA is not available for computations during the reconfiguration,
avoiding some of the reconfigurations can improve overall performance.
This paper presents optimizations based on query sequences,
which reduces the impact of the reconfigurations.
Knowledge of upcoming queries is used to
(I) speculatively start reconfiguration already when a query is still running and
(II) avoid overwriting of reconfigurable regions that will be used again in subsequent queries.
We evaluate our optimizations with a calibrated model and measurements for various parameter values.
Improvements in execution time of up to 28\% can be obtained
even with sequences of only two queries.
\end{abstract}

\maketitle

\section{Introduction}
\label{sec:introduction}

A number of projects are already addressing
the acceleration of database query processing
with the help of FPGAs, e.\,g.\ \cite{Muel09a,Sukh13a,Owai19a},
and their integration into a DBMS.
The ReProVide project \cite{Bech19a} is one of them.
The particular approach of this project is
to use dynamic reconfiguration of the FPGA
with a combination of
a novel DBMS optimizer and
accelerated data-storage units called ReProVide Processing Units (RPUs).
On these RPUs,
a library of query-processing modules is available,
which can be configured onto the FPGA in about 15\,ms.
Due to the limited amount of logic resources on an FPGA,
not all modules can be made available simultaneously.
So at some point in time,
only a subset is ready for use in query processing.
Hence, reconfiguration of one or more region of the FPGA is needed
to process an incoming query optimally,
if the required modules are not loaded already.

The idea presented in this paper is
that when a sequence of queries to be executed repeatedly is known,
information about this sequence can be given to the RPU
via so-called \emph{hints}.
The RPU can use this information
to reduce the overall execution time of the sequence of queries.
Of course, such knowledge about queries could also be used for other purposes,
e.\,g.\ the creation of join indexes or materialized views.
In this paper, however, we assume that the decision to a put some of the data on the RPU
has already been made
(based on an analysis of the workload going well beyond query sequences),
and so the RPU must be involved anyway to get these data.
We will indicate where such sequences of queries can be found.
They are e.\,g.\ part of an application program
that fills the frames of a modular screen output
from different parts of an underlying database.

The paper is structured as follows:
In Section~\ref{sec:reprovide}
we will introduce the RPUs in more detail.
Other approaches to hardware acceleration
and the matching query optimization
will be presented
in Section~\ref{sec:related-work} on related work.
The query sequences are detailed
in Section~\ref{sec:model},
while Section~\ref{sec:exec_model} describes
the execution of queries involving the RPU.
Section~\ref{sec:optimization} addresses the optimization,
and Section~\ref{sec:evaluation} finishes with an evaluation.

\section{The ReProVide Processing Unit (RPU)}
\label{sec:reprovide}

This section gives some more detail on
how the ReProVide system does the query processing.
It is a ``system on chip'' (SoC)
with its own storage (SSD),
an ARM processor,
and some memory (DRAM)---in addition to the FPGA.
So the RPU hosts some datasets (tables),
which means that any access to these tables
must use the RPU\footnote{In this paper, we do not consider the caching of tables in the host system.}.
The RPU is connected to a host via a fast network.
At this host a (relational) DBMS is running,
executing queries that access the tables on the RPU
and combine them with data from other tables stored at the host itself.
It is important to note that
streaming a table through the FPGA
comes at no additional cost\footnote{We assume that the datarate of the network
is smaller than the datarate of the RPU storage.}.

However, reconfiguration may be required
before the RPU can process the data in the way
a particular query demands.
The FPGA of a RPU contains various static hardware modules,
like a storage controller, a network controller, data interconnects, and local memories,
as well as multiple partially reconfigurable regions (PRs) \cite{Bech19a}.
Data is processed by so-called accelerators loaded into these PRs.
RPUs execute a partial query
by streaming the tables from the storage
at line-rate
through one or multiple accelerators
to the network interface.
Operations like sorting or joining of larger tables
cannot be implemented in an efficient way on such a streaming architecture
and are therefore left to the DBMS.
As the accelerators are optimized for line-rate processing,
and FPGA resources are limited,
not all available operator modules can be combined into a single accelerator.
E.\,g.\ implementations of arithmetical operators (mult, add, \ldots )
differ in the data type they operate on (float, int32, int64).
Dynamic partial reconfiguration allows
to offer support for more and more operators
with a growing library of available accelerators,
each implementing a reasonable subset of all available operators.

In the current prototype,
we have started with a set of operators
doing filtering, projection, and semi-join \cite{Bech19a}.
The set is still growing.
We have begun with integers only,
but are now also handling strings and floats.
Furthermore, arithmetic is possible,
so attribute values can be used in calculations,
before they are compared with constants or with each other.
The comparisons are the usual theta operators, i.\,e.\ $<, >, =, \neq, \leq, \geq$.

So the RPU has a state,
which consists of the sequence of accelerators currently configured on the FPGA.
They can be executed immediately.
Processing may even bypass an accelerator,
which remains to be configured.
The other accelerators are available in a library,
but are currently not configured on the FPGA.
In order to execute them,
some additional cost (time) for reconfiguration will be encountered.

A special approach of the ReProVide project is
that the interface of the RPU allows to send some \emph{hints}
accompanying a query-execution request.
These hints do not change any functionality,
but give some information to the RPU
that allows it to optimize the execution further.
This paper underlines the significance of such hints
to avoid unnecessary reconfiguration
or to begin with reconfiguration for upcoming queries
while still executing the current one.

The optimizer of the DBMS that hosts the RPU
identifies the filter operations that can be pushed down to the RPU.
This may greatly reduce the amount of transferred data,
and thus not only relieves the network of unneeded traffic,
but also reduces the load on the DBMS,
as less data must be processed by it.
A cost model can be used
to guide the selection of operations to be pushed down
in case there are not enough PRs available for the given query.

\section{Related Work}
\label{sec:related-work}

Query-sequence optimization has already been addressed in
\cite{Schw01a,Kraf03a} quite a while ago.
They focused on
the optimization of a sequence of SQL statements
generated by a ROLAP query generator.
The coarse-grained optimization (CGO) they introduced
has optimization strategies that include
rewriting the queries
to combine them,
using a common \code{WHERE} clause, \code{HAVING} clause, or \code{SELECT} clause,
and to partition them,
enabling parallel execution.
The decision which optimization strategy to use
is only based on information about the query sequence itself.
I.\,e.\ they started to optimize the sequence
once they have all queries.

Multi-query optimization
can also be used in this regard \cite{Chau16a,Chen98a,Sell88a},
because materialization and reuse of intermediate results
as well as common-sub-expression evaluation
are good choices,
if we have the set of queries all together.
But in our case
the queries will be submitted one after the other, not together.
The execution of a query must be completed
and the result must be returned,
before the next query can be expected.

A query sequence is of interest to us
if it appears frequently in the query log.
The optimization strategies we follow
are based on these patterns that we identify
and the capabilities of the underlying hardware
that processes the queries.
We focus on retrieving the sequence information
when the first query arrives
and utilize the knowledge about the upcoming queries
as hints for preparing the hardware.
Hardware acceleration has not been considered
in any of the mentioned projects.

Query processing using new hardware
has been studied extensively in recent years.
The Hybrid Query Processing Engine (HyPE) \cite{Bres12b,Bres13a,Bres12a}
is a self-tuning optimizer framework.
As the name indicates, it allows for hybrid query processing,
that is, utilizing multiple processing devices.
CoGaDB has used it to build a hardware-oblivious optimizer,
which learns cost models for database operators
and efficiently distributes a workload to available processors.
It is hardware- and algorithm-oblivious,
i.\,e.\ it has only minimal knowledge
of the underlying processors
or the implementation details of operators.
In contrast to our work,
HyPE does not address multi-query optimization and hints.

The ADAMANT project \cite{Bech18b}
enhances a DBMS
with extensible, adaptable support for heterogeneous hardware.
The main objectives are
to find a useful abstraction level
that considers the fundamental differences between the heterogeneous accelerators,
the development of an improved optimization process
that copes with the explosion of search space
with new dimensions of parallelism, and
to fine-tune device-dependent implementation parameters
to exploit the different features available in heterogeneous co-processors.
The project mainly uses OpenCL and emphasizes GPUs,
but is also beginning to look into FPGAs.

Offloading operations for query processing to FPGA-based hardware accelerators
has been researched well
because of its small energy footprint
and fast execution
\cite{Naja13a,Muel09a,Bech15a,Wood14a,Zien16a,Bech16a}.
Our ReProVide system also utilizes these advantages of an FPGA for query processing.
The other work related to FPGA
is all focusing on the execution of a single query.
We extend this with the execution of a sequence of queries
to reduce the hardware-related overhead.

\section{Query-Sequence Model}
\label{sec:model}

The information base that drives the optimization is
a sequence of (relational) queries to be executed repeatedly.
It can be obtained from database query logs \cite{Wahl16a,Schw16a,Wahl18a,Khou09a,Yan18a}
or by code analysis \cite{Smit11a,Nagy13a},
including the frequency and the time of arrival.
The sequence can be adjusted manually
to focus on the most important queries,
if necessary.

We denote a query sequence $S$ containing $n$ queries
as an ordered set of individual queries $Q_0, \dots, Q_{n-1}$.
All queries are run to completion before the next starts
without any overlap in their execution.
The time gap between two successive queries
is estimated from experience
or is learned by repeated observations.
For any query sequence $S$,
the time gap $t_{gap,q}$ with $q \in \{Q_0, \dots, Q_{n-2}\}$ gives
the average time between two consecutive queries,
namely $q$ and its successor.
Please note that the optimization target in this model
is not to optimize the execution time of a single query,
but the time of the whole sequence,
that is, the time from the arrival of $Q_0$
until the transmission of the last result of $Q_{n-1}$,
including all time gaps.

The analysis of the queries in the sequence then
determines
\begin{itemize}
\item the tables and attributes accessed, and in which order, as well as
\item the operators they request on the attributes of the tables
  (arithmetic and/or comparisons)
\end{itemize}
While the proposed approach is suitable for arbitrary operators,
we focus on filters (selections) together with arithmetic for now.
They can easily be extracted
from query-execution plans (QEPs).
The constants used in the arithmetic or in the comparisons
may come from program variables,
so they will be different in future executions of the query sequence.
We will introduce parameters here,
as known from prepared queries and stored procedures.

The query sequences are generated by application programs
that e.\,g.\ compose output from various parts of a database.
Imagine an e-mail client displaying overview lists
together with a preview of the first e-mail on the list
and maybe some attachments.
The first query retrieves the list elements with selected information
such as sender and subject line.
Using the id of the top line,
the second query fetches more details,
e.\,g.\ the list of other recipients and the first 10 lines of the body.
The third query finally obtains previews of the attachments.
One can easily imagine other client interfaces
consisting of frames
filled with related contents
extracted by queries from a database.
A similar, but much more elaborated scenario can found
in the Tableau software accessing Hyper \cite{Voge19a}.

\section{Execution Model}
\label{sec:exec_model}

In order to keep the model simple,
we use only a single PR of the RPU
and do not overlap table scanning with accelerator execution.
We do, however, allow the scanning of tables
with simultaneous accelerator reconfiguration of the PR.

\begin{figure*}[t]
\begin{center}
\includegraphics[width=\textwidth]{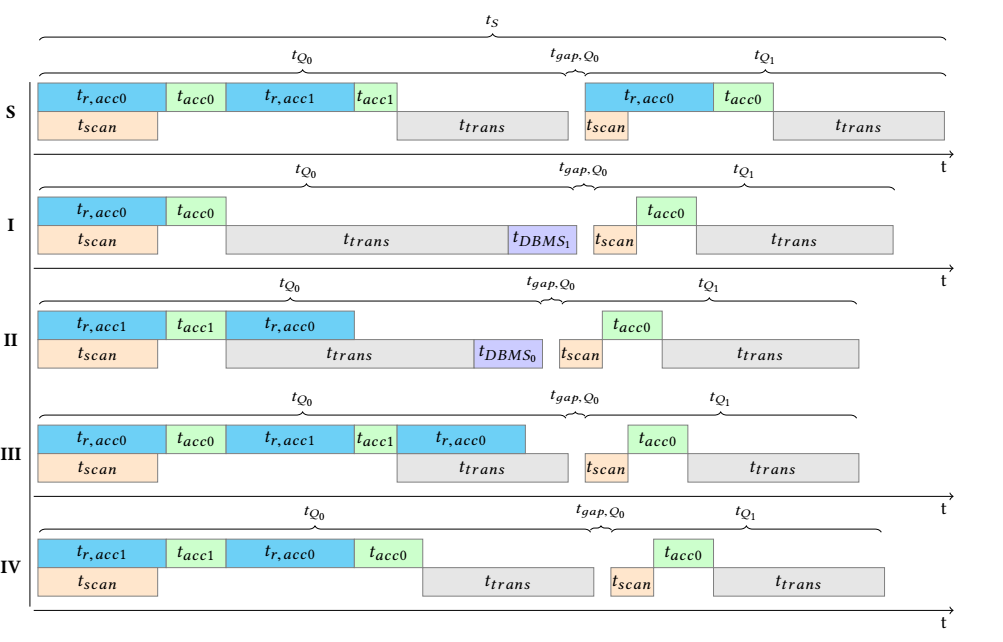}
\end{center}
\caption{
  Bar chart of the execution phases
  in possible global plans
  for a query sequence $S$.
  (S) pushes down both operations of $Q_0$ to the RPU without giving any hints.
  (I) pushes down only the 1st operation,
  so the accelerator configured in the RPU can be reused in $Q_1$
  without reconfiguration.
  (II) pushes down the 2nd operation,
  which may be more selective than the first,
  and then does the required reconfiguration in parallel with
  the data transfer to the host,
  so again the reconfiguration at the beginning of $Q_1$ can be avoided.
  (III) speculatively loads the accelerator for $Q_1$,
  as soon as the 2nd accelerator of $Q_0$ has finished.
  (IV) swaps the accelerator invocations of $Q_0$,
  so that the 2nd accelerator is already loaded for $Q_1$.
}
\label{fig:opts}
\Description{
  A bar chart of the execution phases
  in possible global plans
  for a query sequence $S$
  is shown.
  Plan (S) pushes down both operations of $Q_0$ to the RPU without giving any hints,
  so that at the beginning of $Q_1$ reconfiguration must take place.
  Plan (I) pushes down only the 1st operation,
  so the accelerator configured in the RPU can be reused in $Q_1$
  without reconfiguration.
  Plan (II) pushes down the 2nd operation,
  which may be more selective than the first,
  and then does the required reconfiguration in parallel with
  the data transfer to the host,
  so again the reconfiguration at the beginning of $Q_1$ can be avoided.
  Plan (III) speculatively loads the accelerator for $Q_1$,
  as soon as the 2nd accelerator of $Q_0$ has finished,
  so it may already be available when $Q_1$ starts.
  Plan (IV) swaps the accelerator invocations of $Q_0$,
  so that the 2nd accelerator is already loaded for $Q_1$.
}
\end{figure*}

Fig.~\ref{fig:opts} shows the possible execution plans
of a query sequence $S$ with two consecutive queries ($Q_0, Q_1$).
The general assumption is
that the second query reuses one of the operations of the first query.
This operation can be executed by the same accelerator on the RPU ($acc0$).
This is exactly the information
that we can obtain by analyzing the query logs.
We further assume that the two operations of $Q_0$ commute.
This is a minimum-size sequence of only two queries.
Any other sequence with more queries
(and thus more accelerator runs)
will only enhance the possibility
to apply the optimizations presented here.

The baseline (plan S) is simply executing the query operations in order
and completely on the RPU,
without considering the reconfigurations required for that.
Here $t_{S}$ is the total execution time of the sequence $S$.
The execution time $t_{Q_i}$ of a query
is the sum of the reconfiguration times ($t_{r,accX}$),
the accelerator execution times ($t_{accX}$)
and the network transport time ($t_{trans}$).
As the table scan can run
while the first accelerator is being reconfigured ($t_{r,accX}$),
only the maximum value of $t_{scan}$ and $t_{r,accX}$ is added
to the execution time of the query.

Optimizations I and II in Fig.~\ref{fig:opts} are other possible execution plans,
where the filter operations of the first query are treated differently,
such that only one is pushed down to the RPU
and the other is executed afterwards by the DBMS
(which takes the time of $t_{DBMS_X}$).
In optimization \textbf{I},
the RPU accelerates only the first operation
(using the times $t_{r,acc0}$ for reconfiguration and $t_{acc0}$ for processing)
and then transfers the result to the DBMS,
which executes the second operation ($t_{DBMS_1}$).
This avoids the reconfiguration time for the first operation of query $Q_1$,
but it may result in a larger data transfer.
Optimization \textbf{II} accelerates only the second operation
(which may help to reduce that data transfer, if the selectivity is higher)
and executes the first in the DBMS.
This would require to reconfigure at the beginning of query $Q_1$.
However, knowing the query sequence we can inform the RPU via hints earlier,
so it can do the reconfiguration ($t_{r,acc0}$) in parallel to the data transfer.

Alternatively,
we can still push down both operations to the RPU,
but also give it some hints regarding the query coming next.
This allows the RPU to do some optimizations by its own.
This will be addressed in the following section.

\section{Hardware-Cognitive Optimization}
\label{sec:optimization}

The optimizer of the DBMS
analyzes the sequence of query operations for similarity,
that is, common subexpressions \cite{Chen98a}.
Since the goal is to avoid unnecessary reconfigurations,
only information about subsequences of common comparisons is passed to the RPU as hints.
This information can be organized and extracted
with the help of the query repository \cite{Schw16a}.

There is no scheduling of queries;
a query sequence may begin at any time.
The DBMS optimizer tries to recognize the first query of a sequence
and then retrieves the relevant sequence information,
i.\,e.\ the similarities in the sequence of comparisons.
From that, the optimizer generates the hints to the RPU
in order to avoid reconfiguration
and thus to reduce the execution time
of the pushed-down query plans.

In Fig.~\ref{fig:opts},
the proposed reconfiguration strategies of the RPU are shown
when there is a complete push-down of operations
(again for the plan S).
The RPU can
\begin{itemize}
  \item speculatively load accelerators for subsequent queries,
    once other accelerators are no longer used by the current query (optimization \textbf{III}), and
  \item avoid the replacement of reusable accelerators
    by swapping accelerator invocations (optimization \textbf{IV}).
\end{itemize}
The effect of optimization III is shown Fig.~\ref{fig:opts}.
While the processing of the first query $Q_0$ of the sequence remains unchanged,
the reconfiguration back to accelerator 0 ($acc0$)
is speculatively started
as soon as accelerator 1 ($acc1$) has finished.
In this bar chart,
the reconfiguration---which runs in parallel
with the result transmission---has
completed before the subsequent query $Q_1$ arrives.
This is the optimal case for this optimization.
Query $Q_1$ can start immediately when it arrives,
without any waiting time introduced by reconfiguration.

Optimization IV tries to avoid the third reconfiguration at all
by swapping the accelerator invocations of $Q_0$.
This may not be possible in any case,
but there are cases that allow this.
For instance, two accelerators implementing filters
can be swapped safely.
The local optimizer would always invoke the filter with the lowest selectivity first
to reduce the data volume early.
The swapping, however, may contradict this.
The lower part of Fig.~\ref{fig:opts} shows the consequences:
Query $Q_0$ now needs more time,
as the second accelerator $acc0$ must process more data
and thus takes longer.
While this looks detrimental to the goal of optimization,
the third reconfiguration has in fact vanished
and thus the overall execution time for the complete sequence
is reduced.

Of course this only makes sense,
if it is worthwhile to push down all operations to the RPU
and if the RPU actually reacts to the hints.
That is, the optimization works on the RPU side,
because it skips some time-consuming action
or exploits waiting times.

\section{Evaluation}
\label{sec:evaluation}

We have implemented a ReProVide prototype,
which is demonstrated at the 2020 EDBT conference \cite{Been20a}.
The host is running Apache Calcite \cite{Bego18a} as a DBMS.
It has been extended with optimization rules
to push down available operators to RPU based on cost.
With this prototype,
we have conducted a series of measurements
providing us with numbers for the time variables
used in the bars of the diagram of the query sequence $S$
presented in Fig.~\ref{fig:opts}.

We have evaluated all the optimizations
shown in Fig.~\ref{fig:opts}
by varying the table size $s_{table}$
to be processed by the two queries
(here we considered the \code{date\_dim} table from the TPC-DS benchmark)
from initially 9\,MB for the first query
and from an initial table size of 1\,MB for the second query
with a scale factor from 1 to 5.
We have measured the data rate of the SSD scan
to be $r_{scan} =$ 1\,GB/s,
so we can calculate
\begin{equation}
t_{Q_i,scan} = s_{Q_i,table} \times r_{scan}
\end{equation}
The reconfiguration time $t_{r,accX}$ is known to be 15\,ms
for each accelerator.
The data rate $r_{acc}$ of the accelerators is 1.5\,GB/s,
so that
\begin{equation}
t_{Q_0,acc0} = s_{Q_0,table} \times r_{acc}
\end{equation}
Currently, all our accelerators are filters,
so they reduce the size of the table
by a selectivity factor of $f_{Q_i,accX}$.
This leads to an intermediate table of size
\begin{equation}
s_{Q_0,intermediate} = s_{Q_0,table} \times f_{Q_0,acc0}
\end{equation}
and a final result size of
\begin{equation}
s_{Q_0,result} = s_{Q_0,intermediate} \times f_{Q_0,acc1}
\end{equation}
So for query $Q_1$ we get
\begin{equation}
s_{Q_1,result} = s_{Q_1,table} \times f{Q_1,acc0}
\end{equation}
The result is then transferred to the host
at a network data rate of $r_{network}$ = 80\,MB/s.
\begin{equation}
t_{Q_i,trans} = s_{Q_i,result} \times r_{network}
\end{equation}
Overall, we get an execution time for the unoptimized plan for query sequence $S$ of:
\begin{multline}
t_{Q_0} = max(t_{r,acc0}, t_{Q_0,scan}) + t_{Q_0,acc0} + t_{r,acc1} \\
            + t_{Q_0,acc1} + t_{Q_0,trans}
\end{multline}
\begin{equation}
t_{Q_1} = max(t_{r,acc0}, t_{Q_1,scan}) + t_{Q_1,acc0} + t_{Q_1,trans}
\end{equation}
\begin{equation}
t_S = t_{Q_0} + t_{gap,Q_0} + t_{Q_1}
\end{equation}
For the optimization I in Fig.~\ref{fig:opts},
the calculation changes to:
\begin{multline}
t_{Q_0} = max(t_{r,acc0}, t_{Q_0,scan}) + t_{Q_0,acc0} + t_{Q_0,trans} \\
          + t_{Q_0,DBMS1}
\end{multline}
with $t_{Q_0,DBMS_1}$ as the time for the DBMS operator execution.
We have measured that
and have found it to be approx.\ 0.03\,ms for each MB of data.
\begin{equation}
t_{Q_1} = t_{Q_1,scan} + t_{Q_1,acc0} + t_{Q_1,trans}
\end{equation}
\begin{equation}
t_S = t_{Q_0} + t_{gap,Q_0} + t_{Q_1}
\end{equation}
For the optimization II in Fig.~\ref{fig:opts},
the calculation changes to:
\begin{multline}
t_S = max(t_{r,acc1}, t_{Q_0,scan}) + t_{Q_0,acc1} \\
        + max(t_{r,acc0}, (t_{Q_0,trans} + t_{Q_0,DBMS0} + t_{gap,Q_0})) + t_{Q_1}
\end{multline}

In optimization I,
we save the reconfiguration time in $Q_1$,
but need more transfer time and also DBMS postprocessing in $Q_0$.
For smaller selectivities $f_{Q_0,acc0}$
and smaller tables it is still better than S.

\begin{figure}[H]
\includegraphics[width=\columnwidth, height=.3\textheight]{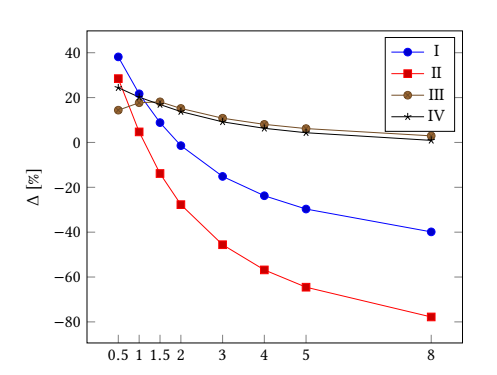}
\caption{Execution-time improvement over relation-size scaling for optimizations I and II
}
\label{fig:execVsglobalopt}
\Description{The graph shows the execution-time improvement over the relation-size scaling for the optimization plans PI and PII.
  It can be seen that there is no improvement at all, to the contrary, for all scale factors above 1 the improvement is negative.
}
\end{figure}

The percentage of improvement in total execution time ($t_S$)
for I and II over the standard plan S is shown in
Fig.~\ref{fig:execVsglobalopt}.
Here, the time gap has been set to 1\,ms,
and the selectivities are 33\% for the first operator of $Q_0$,
43\% for the second,
and 14\% first the only operator of $Q_1$.
The negative improvements with relation size scaling
make clear that
the complete push-down of operations is preferable here
(except for very small relations)
because of the huge data-transfer overhead.
However,
this will be different for stronger selectivities.

This results suggest that
in a scenario like this,
the complete push-down of all operations and
the hardware-cognitive optimization described in Section~\ref{sec:optimization}
is preferable.
The calculations
to evaluate the performance of the optimizations III and IV
are given below.

For optimization III,
the total time changes to:
\begin{multline}
t_S = max(t_{r,acc0}, t_{Q_0,scan}) + t_{Q_0,acc0} + t_{r,acc1} + t_{Q_0,acc1} \\
        + max(t_{r,acc0}, (t_{Q_0,trans} + t_{gap,Q_0} + t_{Q_1,scan})) \\
        + t_{Q_1,acc0} + t_{Q_1,trans}
\end{multline}
For optimization IV,
the calculations look like this:
\begin{multline}
t_{Q_0} = max(t_{r,acc1}, t_{Q_0,scan}) + t_{Q_0,acc1} + t_{r,acc0} \\
          + t_{Q_0,acc0} + t_{Q_0,trans}
\end{multline}
\begin{equation}
t_{Q_1} = t_{Q_1,scan} + t_{Q_1,acc0} + t_{Q_1,trans}
\end{equation}
\begin{equation}
t_S = t_{Q_0} + t_{gap,Q_0} + t_{Q_1}
\end{equation}

\begin{figure*}[ht]
	\begin{subfigure}[t]{.33\textwidth}
%
%
%
%
%
       \includegraphics[width=\columnwidth, height=.2\textheight]{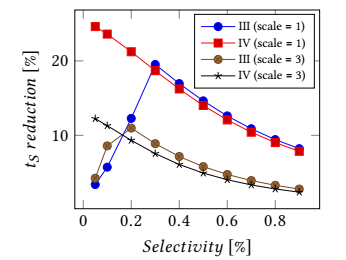}
		\caption{Over selectivity for the two relation-size \\ scale factors 1 and 3 and a $t_{gap}$ of 1\,ms}
		\label{fig:execVsSelectivity}
		\Description{The execution-time improvement over the selectivity
			is approximately 30\% for small selectivity and small relations
			and goes down to 5\% for larger selectivity and larger relations.}
	\end{subfigure}
	\begin{subfigure}[t]{.33\textwidth}
%
%
%
%
%
       \includegraphics[width=\columnwidth, height=.2\textheight]{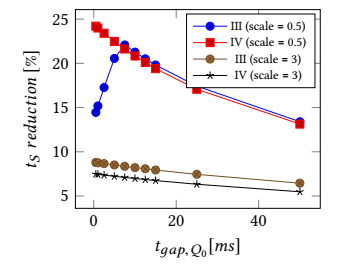}
		\caption{Over $t_{gap}$ for the two relation-size \\ scale factors 0.5 and 3 and the \\
      standard selectivity of 33\%}
		\label{fig:execVstgap}
		\Description{The execution-time improvement over the average time gap between two queries
			is approximately 25\% for smaller gaps and goes down to 5\% for larger gaps and larger relations.}
	\end{subfigure}
	\begin{subfigure}[t]{.33\textwidth}
%
%
%
%
%
      \includegraphics[width=\columnwidth, height=.2\textheight]{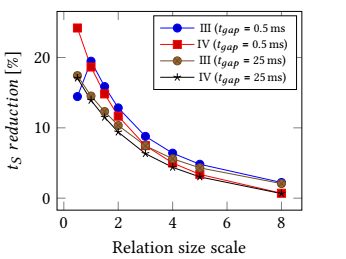}
		\caption{Over relation-size scale factors for \\ the two gap sizes of 0.5\,ms ($t_{gap} \ll t_r $) and 25\,ms ($t_{gap} \gg t_r$)
      with selectivity 33\%}
		\label{fig:execVsScale}
		\Description{The execution-time improvement over the average time gap between two queries
			is above 20\% for smaller gaps and goes down to 5\% for larger gaps and larger relations.}
	\end{subfigure}
	
	\caption{Execution-time improvement compared to S for optimizations III and IV}
	\label{fig:evaluationResults}
	
\end{figure*}

Fig.~\ref{fig:evaluationResults} shows the results of these calculations
with respect to various factors.
The main factors that influence the execution time
of a sequence of queries are:
selectivity ($f_{Q_i,accX}$),
time gap between the queries ($t_{gap,Q_0}$), and
size of the table to be processed ($s_{Q_i,table}$).
The performance improvements
of the suggested optimizations III and IV over these factors
and with respect to S have been calculated
using the formulae given above.
Fig.~\ref{fig:execVsSelectivity} shows that
for small selectivities (with a fixed $t_{gap}$ of 1\,ms $\ll t_{r,accX}$)
optimization IV is the best.
If only a small amount of data is to be transferred after filtering,
then performing additional reconfiguration generates an overhead in optimization III.
This leaves an execution-time improvement of up to 28\%
for optimization IV.
A $t_{gap}$ of 25\,ms $\gg t_{r,accX}$ avoids this overhead in optimization III,
which is shown in Fig.~\ref{fig:execVsScale}, 
where the selectivity is fixed
(at the three numbers given above, namely 33\%, 43\%, and 14\%).
Here, two different time gaps between the queries are used,
namely 0.5\,ms ($t_{gap} \ll t_r$) and 25\,ms ($t_{gap} \gg t_r$).
If there is a considerable time gap between the queries in the sequence,
the RPU can prepare the hardware for the upcoming query during this time and
hence the optimization III will perform better,
as can be seen in Fig.~\ref{fig:execVstgap}.
Here again the selectivity is fixed for the two queries
at the three mentioned values.
Fig.~\ref{fig:execVsScale} finally indicates that
for small relation sizes and a small time gap,
optimization IV is preferable
and can lead to an execution-time improvement of up to 24\%.

It can be clearly seen that
optimization IV drastically outperforms optimization III
for small relation sizes and for small selectivities,
where $t_{trans} + t_{gap,Q_0} + t_{scan,Q_1} \leq t_{r,acc0}$.
Optimization III dominates when this breaks,
because the reconfiguration time can be hidden completely from that point on.
While the result sizes vary with the parameter values in the queries,
the time gap $t_{gap,Q_0}$ obtained by the sequence analysis
gives a good indication
of the case where the speculative reconfiguration approach should be used
and how effectively the reconfiguration time can be hidden from the total execution time.

\section{Conclusion}


The paper has introduced the utilization of information on query sequences
in the optimization of processing them on reconfigurable accelerator hardware.
The ReProVide Processing Unit (RPU)
is such a system
that can filter data on their way from storage to DBMS.
The reconfiguration required to adapt to the next query
can be substantially reduced
by taking the sequence of coming queries into account.
Avoiding reconfiguration times
by not pushing down filter operations to the RPU
turned out to have benefit only
in the case of rather small table sizes.
So the choice is instead
to push down all filter operations
and give the RPU some hints
so it can do some local optimizations.
It can (I) reconfigure in parallel to the result-data transfer of a running query
and can (II) be avoided reconfiguration completely by swapping accelerators.
The second optimization has strong improvement effects for smaller relations
and for smaller selectivities,
while the first overtakes for larger relations and/or larger time gaps between queries.

The optimizations done by hand in these evaluations
already show promising results.
So we will be built them into the optimizer.


For further optimizations,
we will use query-analysis graphs similar to those proposed in \cite{Chen98a}.
The idea is to keep result data in the memory of the RPU,
if they can be reused in subsequent queries
for common sub-expressions and subsumption.

More hints can be imagined
to optimize the RPU further:
We could synthesize other accelerators
by combining various arithmetic and comparison operators,
depending on the frequency of these combinations in the query sequences.
Also, for higher table scale factors, 
the RPU will rather provide an index lookup instead of a full table scan. 
Which index to create can be derived from the query sequences.
This would reduce $t_scan$ significantly.

\subsubsection*{Acknowledgements}

This work has been supported by the German Science Foundation
(Deutsche Forschungsgemeinschaft, DFG)
with the grant no.\ ME~943/9-1.

\bibliographystyle{ACM-Reference-Format}
\bibliography{paper03}

\end{document}